\documentclass[11pt]{article}
\usepackage{amssymb,amsmath,cite,geometry,url}
\catcode`\@=11

\@addtoreset{equation}{section}
\def\theequation{\arabic{section}.\arabic{equation}}
     
\catcode`\@=11
\def\thesection{\arabic{section}}

\def\appendix{\setcounter{section}{0}
        \def\thesection{Appendix.}
        \def\theequation{\Alph{section}.\arabic{equation}}}
\def\section{\@startsection{section}{1}{\z@}{3.5ex plus 1ex minus
   .2ex}{2.3ex plus .2ex}{\large\bf}}
\def\subsection{\@startsection{subsection}{1}{\z@}{3.5ex plus 1ex minus
   .2ex}{2.3ex plus .2ex}{\bf}}
     
\long\def\@makefntext#1{\parindent 0cm\noindent
\hbox to 1em{\hss$^{\@thefnmark}$}#1}

\newcommand{\captionfonts}{\small}
\makeatletter  
\long\def\@makecaption#1#2{%
  \vskip\abovecaptionskip
  \sbox\@tempboxa{{\captionfonts #1: #2}}%
  \ifdim \wd\@tempboxa >\hsize
    {\captionfonts #1: #2\par}
  \else
    \hbox to\hsize{\hfil\box\@tempboxa\hfil}%
  \fi
  \vskip\belowcaptionskip}
\makeatother   

\begin{document}
\begin{titlepage}
\vspace{.5in}
\begin{flushright}
\today\\  
\end{flushright}
\vspace{1.5in}
\begin{center}
{\Large\bf
 Challenges for Emergent Gravity}\\  
\vspace{.4in}
{S.~C{\sc arlip}\footnote{\it email: carlip@physics.ucdavis.edu}\\
       {\small\it Department of Physics}\\
       {\small\it University of California}\\
       {\small\it Davis, CA 95616}\\{\small\it USA}}
\end{center}

\vspace{.5in}
\begin{center}
{\large\bf Abstract}
\end{center}
\begin{center}
\begin{minipage}{4.75in}
{\small
The idea of gravity as an ``emergent'' phenomenon has gained popularity 
in recent years.  I discuss some of the obstacles that any such model must
overcome in order to agree with the observational underpinnings of
general relativity.
}
\end{minipage}
\end{center}
\end{titlepage}
\addtocounter{footnote}{-1}

Over the past few years, the idea of gravity as an ``emergent''
phenomenon has become increasingly popular.  Emergence is seldom 
sharply defined, and the notion has been described as ``vague and 
contentious'' \cite{Butterfield}, but the basic picture is that gravity, 
and perhaps space or spacetime themselves, are collective manifestations
of very different underlying degrees of freedom.  Such proposals are 
not new---Wheeler wrote about pregeometry as early as 1963 
\cite{Wheelera}, and Finkelstein introduced a version of causal set 
theory in 1969 \cite{Finkelstein}---but historically they were marginal 
aspects of research in quantum gravity, whose most noted proponents 
were often outsiders to the field (e.g., \cite{Pines,Laughlin, Volovik}).  
More recently, though, such mainstream approaches as string theory 
matrix models \cite{Seiberg}, the AdS/CFT correspondence 
\cite{Polchinski,Witten}, loop quantum cosmology \cite{Bojowald}, and 
quantum Regge calculus \cite{Loll} have all been described as 
ß``emergent.''

Given the enormous difficulty of quantizing general relativity, it is 
natural to consider the possibility  that we have simply been trying to 
quantize the wrong degrees of freedom \cite{Hu2,BVL}.  But general 
relativity has an enormous body of observational support, and models 
of emergence face formidable challenges in reproducing these successes.  
The goal of this article is to explain some of these challenges.  

This is by no means a comprehensive survey of emergent gravity.  
The specific models I refer to have been chosen primarily as exemplars 
for particular issues, and most of the ideas are well known to those 
working in the field, although they have not been collected in one 
place.  For broader reviews, see \cite{Sindonirev} or articles in 
\cite{Oritibk}.
 
I should also start with a cautionary note.  In one sense, \emph{any}
quantum theory of general relativity will have ``emergent''  aspects.
All observables (in Dirac's sense) are necessarily nonlocal  
\cite{Torre,Hartle}, so locality must be an emergent property.  
Moreover, for asymptotically anti-de Sitter spacetimes, and perhaps for
asymptotically flat spacetimes as well, the algebra of observables is 
isomorphic to an algebra of operators at an asymptotic boundary 
\cite{Marolf1}, so the bulk physics can itself be viewed as emergent 
from a boundary theory.  Such a picture seems contrary to the spirit of 
emergence, since the boundary degrees of freedom are merely 
asymptotic values of ordinary bulk degrees of freedom, but I do not 
know how to make this distinction precise.

\section{Classifying emergent models \label{cats}}

My aim is to describe obstacles facing all models of emergent gravity,
rather than focusing on details of any particular approach.  Still,
it will be useful to distinguish two broad categories of models
that face rather different challenges.

The first category, which I will call type I, comprises models in which
the fundamental degrees of freedom live in some sort of ``environment'': 
a medium, a lattice, a pre-existing space or spacetime, or the like.
Examples include analog models based on fluid flows and similar
phenomena \cite{LV,Unruhb,Wein,Hu},  gravity-like excitations near the 
Fermi point \cite{Volovik,Volovik2}, quantum Hall effect edge states
\cite{Zhang}, deformations in an elastic solid
\cite{Zaanen}, and spin systems on a fixed lattice \cite{Wen,RD}.  
The background environment can have fairly minimal structure;  in 
``pregeometric'' models, for example, composite gravitons appear 
in what is initially a purely topological manifold \cite{Akama,Amati,%
Floreanini,Wetterich,Alfaro}.  Type I models must typically 
decouple the environment from observable quantities quite strongly 
to reproduce observation.

The second category, which I will call type II, comprises models in which
space or spacetime are themselves emergent.  Examples include graph-based
models such as quantum graphity \cite{Marko} and DUCTs \cite{Wall},
group field theory \cite{Oriti,Oriti2}, certain matrix models 
\cite{Steinacker}, Wheeler's notion of ``it from bit'' \cite{Wheeler2}, 
and other attempts to model the Universe as a quantum computer 
\cite{Lloyd,Marko2}.   Arguably, some versions of the AdS/CFT 
correspondence \cite{Polchinski,Berenstein,Murugan}---those in 
which the conformal field theory is considered primary and the bulk 
spacetime emergent---fall into this category; the CFT lives in some 
lower dimensional, nondynamical space, but points in that 
space need not have any relationship with points in our bulk spacetime,
so the latter might be considered emergent.  In type II models, before 
even asking about contact with observation, one must typically work 
quite hard to find observables that display the large scale existence 
of spacetime at all.

The distinction I am making is similar, although not quite identical, to
Bain's differentiation between emergence from a ``spatiotemporal structure'' 
and a ``non-spatiotemporal reality''\cite{Bain}.  It is not a sharp one.  
Causal set models \cite{Bombelli,Sorkin}, for instance, contain elements of 
a pre-existing spacetime in the form of points and their causal relations,
but the continuum is emergent.  If the conformal field theory is viewed as
primary in the AdS/CFT correspondence, our bulk spacetime might be 
considered emergent, but if the correspondence is a true duality, the
bulk spacetime is present from the start.  The hypothesis of 
spontaneous dimensional reduction at short distances \cite{Carlip2,Carlip3}, 
the dynamical generation of extra dimensions \cite{Arkani}, and some 
``CFT primary'' forms of the AdS/CFT correspondence \cite{Murugan}
suggest another possibility: some dimensions may be emergent while 
others are not.   

I am omitting a third category, models in which spacetime,
the metric, and diffeomorphism invariance are present,
but the \emph{dynamics} of gravity is emergent.  The archetype  
is Sakharov's induced gravity \cite{Sakharov,Adler}, in which the 
Einstein-Hilbert action first appears as a counterterm 
in the matter action.\footnote{In some Sakharov-inspired models, 
the metric appears only as a collective field.  I consider such 
models to be emergent; the examples I am aware of are type I.  Note 
that because the metric is a functional of other fields, the resulting 
field equations may differ from those of general relativity even if 
an Einstein-Hilbert action is induced \cite{BVL}.}
From the point of view of effective field theory \cite{Burgess}, the 
distinction between such models and general relativity is marginal: 
an effective action includes all possible terms, and it 
is not clear that one can distinguish their origins.  Also in this 
category are models in which the dynamics of the metric is determined by 
thermodynamics \cite{Jacobson,Padmanabhan,Pad1}, or as a consistency 
condition for the propagation of other fields \cite{Pad2}.  While these 
are certainly interesting, they fail my (perhaps narrow) criteria for 
emergence unless the metric is absent as a fundamental field.

\section{Gravity as a metric theory \label{EPSsec}}

Before proceeding further, it is worthwhile to review the physical
basis for treating gravity as a metric theory.  As first discussed by
Weyl and later developed by Ehlers, Pirani, and Schild
\cite{EPS}, the observations of local Lorentz invariance and the
universality of free fall allow one to construct a metric description
of gravity.   Briefly summarized, the argument goes as follows
\cite{EPS,Woodhouse} (for a somewhat different approach, see 
section 2.3 of \cite{Will}):
\begin{enumerate} 
\item {\bf Local Lorentz invariance} implies the existence of a field 
of light cones, which establishes a causal structure and a topology.
The light cones also determine a conformal structure---an equivalence 
class of metrics that differ only by local rescalings---for which paths 
of light rays are null geodesics. 
\item {\bf The equivalence principle}, specifically the universality 
of free fall, determines a set of preferred paths in spacetime, the 
trajectories of freely falling structureless objects.  Such a projective 
structure, in turn, determines an equivalence class of affine connections 
for which these paths are geodesics.  
\item {\bf Compatibility} of these two structures---the observation
that the trajectories of freely falling massive objects lie within the
light cones but can ``chase'' light arbitrarily closely---fixes a Weyl 
structure, an equivalence class of conformal metrics and affine 
connections such that 
\begin{align}
\nabla_a g_{bc} = A_a g_{bc} \qquad
\text{for some vector field $A$.}
\label{a1}
\end{align} 
\end{enumerate}

As Einstein first noted, though (see \cite{Einstein}), unless $A_a$ 
is of the form $\partial_a\varphi$ for some scalar $\varphi$, eqn.\
(\ref{a1}) implies that lengths change under parallel transport.
This would lead to a ``second clock effect,'' a dependence of the
rate of a clock on its history.  The observation that this is not the 
case in our Universe suggests a further condition:
\begin{enumerate}{\setcounter{enumi}{3}}
\item {\bf The absence of a ``second clock effect''} implies that the 
vector $A$ can be eliminated by a Weyl transformation, picking 
out a unique representative of the conformal class of metrics. 
This requirement can be replaced by a number of others: for 
instance, that neighboring clocks remain synchronized \cite{EPS} 
or that matter waves follow geodesics in the short wavelength
 limit \cite{Audretsch}.
\end{enumerate}

Together, these observations imply that motion in a gravitational field
can be described as geodesic motion in a Lorentzian spacetime, with 
a metric providing a full description of the field.  To go beyond this
kinematic setting and obtain the dynamics of general relativity,
more is needed.  One avenue---certainly not the only one---is this:
\begin{enumerate}{\setcounter{enumi}{4}}
\item {\bf The absence of nondynamical background structures}
implies ``general covariance'' in the sense commonly used
by physicists \cite{Anderson}: gravity should be described by 
diffeomorphism-invariant expressions involving only the metric
and other dynamical fields.  This is a new assumption: while 
Lorentz invariance restricts nondynamical objects, it does not
eliminate them.  A flat background metric, for instance, is
allowed by Lorentz invariance, but permits Nordstr{\o}m's
conformally flat theory and Rosen's bimetric theory (see\cite{Will});
a nondynamical volume element, also allowed by Lorentz 
invariance, permits unimodular gravity \cite{Unruh}.
\end{enumerate}

We must still address the possibility of additional dynamical fields.  
We can, of course, eliminate these by fiat---in a scalar-tensor theory, 
for instance, we can call the purely metric piece of the interaction 
``gravity'' and relegate the scalar to the status of an extra ``fifth 
force.''  Alternatively, note that our kinematic assumptions led to 
a picture in which the response of matter to the gravitational 
field depended solely on the metric.  We might ask that the reciprocal 
response of the gravitational field to matter also occur solely through 
the metric, with no added fields whose sole function is to mediate
between the two (as in, for instance, TeVeS \cite{Bekenstein}).  That is,
\begin{enumerate}{\setcounter{enumi}{5}}
\item {\bf The decoupling of any nonmetric degrees of freedom} from
the dynamics of gravity implies that the gravitational effective action 
should depend on the metric alone.
\item {\bf The methods of effective field theory} then tell us how to 
formulate the action \cite{Burgess}.  At the scales at which a metric 
description applies, the effective action will include all possible 
local, diffeomorphism-invariant functions of the metric that are not 
excluded by other symmetries.  At low energies, this effective action 
will coincide with the Einstein-Hilbert action, albeit with a cosmological 
constant whose magnitude remains a mystery.  One caveat remains: 
the derivative expansion that defines the effective action works only
if the metric varies slowly compared to the Planck scale.  This is 
certainly true observationally, but at a deeper level it is a further
mystery.
\end{enumerate}
 
These arguments do not require that the metric be a
\emph{fundamental} degree of freedom.  Rather, they describe the
setting in which an effective metric description might naturally
emerge.  On the one hand, this provides hope for emergent
gravity: only a few steps are needed to obtain a model that agrees 
with observation.  On the other hand, though, none of these steps
can be avoided, and this places severe restrictions on such models.  
Most of the pitfalls I describe below arise, directly or indirectly, from 
these requirements.

\section{Challenges for models of emergent gravity}

The preceding section established a set of conditions that a model of 
emergent gravity should meet to reproduce general relativity.  While 
such a criterion may be too strong at cosmological or sub-millimeter 
distances, general relativity is extremely well established at intermediate 
scales \cite{Will,Willb}, and any disagreements are strongly constrained 
by observation.  Let us now try to understand the extent to which these
conditions create obstacles for emergent gravity.

\subsection{Lorentz invariance}

We start with Type I models, in which the fundamental degrees of freedom 
live in a space or spacetime environment.  The basic problem 
is clear: unless that background is itself Lorentz invariant, it must decouple 
from observable quantities strongly enough to match observations.  Since
experimental limits on Lorentz violation are very strong \cite{Mattingly}, 
this is a severe restriction.

Perhaps surprisingly, we have simple examples in which such a 
decoupling occurs, at least to lowest order.  In many analog models---%
models in which curved spacetimes are mimicked by phenomena such as 
fluid flows---small perturbations in the flow satisfy Lorentz invariant 
equations, with an effective ``speed of light'' determined by properties 
of the medium \cite{Unruhb,LV}.  Quite generally, linearization 
of a field theory around a nontrivial background leads to an effective 
Lorentzian metric \cite{BarLibVis}, whose signature is fixed by the
hyperbolicity of the partial differential equations.  At higher orders, 
Lorentz violations reappear \cite{Liberati,Lin}, and can serve as 
constraints on such models.

Problems arise, though, as soon as more than one kind of excitation  
can occur.  In that case, distinct excitations typically have distinct 
Lorentz invariances, with different effective metrics and speeds of light 
\cite{BarLibVisb,Sindonirev}.  Even this is a special case; in general, the 
Lorentzian geometry becomes Finslerian, losing contact with the desired 
physics.  One can sometimes recover a single metric by imposing a discrete 
symmetry on the fundamental fields \cite{Sindonib,Girellib}, but this 
step seems rather artificial.  Note that it is \emph{not} enough to 
simply claim that the excitations are all perturbations of the same 
medium: even in an ordinary elastic solid, longitudinal and transverse 
waves travel at different speeds.

One solution is to postulate that the ``environment'' is itself Lorentz 
invariant.  If this environment is a spacetime, this begs the question: 
as in the EPS construction of section 2, invariance implies the existence 
of a conformal class of metrics, and one must explain why these are not 
already dynamical.  Models such as \cite{Wetterich2} in which the 
background has no metric avoid this problem, and  Lorentz invariance may 
be inserted by hand as a gauge symmetry.  Relating this symmetry to 
spacetime Lorentz invariance requires a fairly elaborate scenario, though:
one must ensure the emergence of  a nondegenerate soldering 
form, a tetrad whose two-index structure ``solders'' the fibers in which 
the gauge group acts to the spacetime \cite{Percacci}.  

If the background is discrete, one must work harder.   As Dowker has  
emphasized \cite{Dowker}, Lorentz invariance requires a radical 
nonlocality on a lattice, in the sense that each point has infinitely many 
nearest neighbors \cite{Moore,Bombelli}.  Causal set theories 
\cite{Sorkin,Dowker,Bombelli2} can achieve statistical Lorentz invariance 
with a suitable ``sprinkling'' of spacetime points, but this becomes
much harder if one starts with a more complex discrete structure.  
Causal dynamical triangulations \cite{Loll} may recover Lorentz 
invariance as an average over noninvariant simplicial complexes, but 
this is not certain; the continuum limit of this model may be a
Ho{\v r}ava-Lifshitz theory with a preferred time slicing  
\cite{HLCDT,HLCDT2}.   

One may look instead for models in which Lorentz invariance is 
only recovered at large distances.  After all, ordinary lattice 
quantum field theory is constructed on a lattice that is not even 
rotationally invariant, but by tuning parameters to a second 
order phase transition one can send correlation lengths to infinity, 
wiping out any memory of the underlying lattice.  It is not clear that 
a similar procedure exists for Lorentz invariance, though, where the 
correlations must respect the light cone structure.  One possible
ingredient could be an emergent supersymmetry, which can 
suppress Lorentz violation \cite{Nibbelink}.  

Alternatively, 
Lorentz invariance could appear as a low energy symmetry under the 
renormalization group flow \cite{Nielsen}.  Models are known in 
which different ``speeds of light'' flow to a single value at an 
infrared fixed point \cite{Anber}, but this flow is typically only 
logarithmic in energy, requiring enormous initial energy scales 
or delicate fine tuning to meet observational constraints. 

For type II models, Lorentz invariance may be less contrived. 
By a century-old argument \cite{Baccetti}, the existence of
inertial frames, isotropy of space, and the relativity principle are 
enough to imply Lorentz transformations with some (perhaps infinite) 
``speed of light.''  For type I models, the presence of a background 
typically violates either isotropy or relativity, but this need not
be the case for for type II models.  The effective speed of light might
then be determined dynamically, for example from the Lieb-Robinson 
limit on the speed of information propagation \cite{Marko3} or from
a group structure already present in the fundamental degrees of 
freedom \cite{Girelli}.  But while isotropy may be natural, the relativity 
principle is more problematic.  The Lorentz group is noncompact, 
so transformations must relate inertial frames that are arbitrarily 
``distant'' (although it is again conceivable that an
emergent compact supersymmetry could help).  This 
noncompactness also makes it difficult to achieve Lorentz invariance
by averaging over noninvariant configurations, since the integral
over boosts diverges, though there has been some work on defining
a group average \cite{Marolf}.  

Of course, we have not experimentally tested Lorentz invariance up 
to infinite boost.  But even violations at very high energies can 
feed back into quantum field theory through loop effects and 
lead to drastic consequences at low energies \cite{Collins}, although 
there are proposals for avoiding this problem (e.g., \cite{Pospelov}). 
Small violations of Lorentz invariance also lead to problems with 
black hole thermodynamics: unless black holes simply do not exist in 
the underlying theory, such effects generically violate the generalized 
second law of thermodynamics, allowing perpetual motion machines 
\cite{Dubovsky,Eling}.

\subsection{Principle of equivalence \label{equiv}}

The principle of equivalence takes a number of different guises, not 
all exactly equivalent.  For our purposes, the most relevant version 
is the universality of free fall, with its implication that all forms of 
matter couple to gravity with equal strength.  As Feynman emphasized
\cite{Feynman}, this universality implies a spin two graviton: 
energy falls with the same acceleration as mass, and the unique Lorentz 
covariant combination of mass and energy density, the stress-energy 
tensor, couples to spin two.  A model of emergent gravity must thus 
ensure that 
\begin{enumerate}
\item only one massless spin two field is relevant;
\item this field couples with equal strength to all matter;
\item any spin zero or spin one components of the interaction
are absent or strongly suppressed.  (In some models \cite{Zhang},
higher spin interactions must also be suppressed.)
\end{enumerate}
The principle of equivalence is extremely well tested from millimeter to 
Solar System distances \cite{Will,EotWash}, so while the very short 
and long distance behavior may differ, these requirements 
are quite strong.

In type I models, the would-be gravitational degrees of freedom typically
have no initial connection to the geometry; their role as a metric
emerges later.  Hence there is no obvious reason to expect only a single 
massless spin two field.  For a large class of models built from field 
theory fluctuations around a linearized background \cite{BarLibVisb},
for instance, many ``gravitons'' appear, and the imposition of an ad hoc 
symmetry \cite{Sindonib} is the only known way to force universality. 
In models of composite gravitons, a similar multiplicity of potential 
metrics occurs \cite{Wetterich4}.  One can argue that if the 
effective action is invariant under diffeomorphisms and local Lorentz 
transformations, only one such field will remain massless, 
with the others acquiring large masses \cite{Wetterich4}---typically 
on the order of the Planck mass, although approximate symmetries can 
make them smaller \cite{Damour2}.  The requirement of exact invariance 
is, of course, a very strong one: strong enough, in fact, to \emph{forbid} 
more than one massless graviton \cite{Boulanger}.  But this solution is 
also problematic, since most models with massive spin two fields are 
sick, containing negative energy Boulware-Deser ghosts \cite{BD}.  
While a few exceptions exist \cite{Hinterbichler,deRham,Hassan,RRosen}, 
these require a very special form of the action, and it is not at all clear 
how such a feature would emerge from a more primitive model.

Once one has a single metric, though, a result of Weinberg offers a path 
for deriving the equivalence principle \cite{Weinberg}.  The ``soft 
graviton theorem'' shows that a Lorentz-invariant, massless spin two 
particle that can scatter nontrivially must couple universally to a
single conserved stress-energy tensor.  One must be careful of 
assumptions here; see the discussion below of the Weinberg-Witten 
theorem.  In particular, Weinberg's result are only relevant if
Lorentz invariance has already emerged.  But the theorem suggests
that if only one metric is present, universal coupling may not be  
arbitrary, but may be associated with the universality of Lorentz  
invariance.

For type II models, the primary question comes earlier: does a dynamical 
spacetime emerge at all?   If the fundamental degrees of freedom generate 
such a spacetime at some scale, a metric description offers a natural 
way to describe the dynamics.  As in ordinary general relativity, one 
might then expect a single spacetime to have a single metric.  On the 
other hand, if matter emerges from the same degrees of freedom, 
Weinberg's soft graviton theorem, with its requirement of Lorentz 
invariance, is the only reason I know to expect universal coupling. 
Since very few type II models are yet able to describe the
coupling of matter to gravity, much less to compare couplings of more
than one species of matter, the problem remains almost completely open.

\subsection{Self-coupling \label{self}}

One aspect of universal coupling deserves special attention: we observe 
gravity's coupling to its own energy to occur at the same universal 
strength as its coupling to matter \cite{Will,LLR}.  This self-coupling 
implies that the interaction is nonlinear, and, in fact, it can be used to 
determine the nonlinear terms, giving another route to the Einstein field 
equations \cite{Feynman,Kraichnan,Deser,Deser2}.

This property places requirements on emergent models  beyond 
the linear approximation.  Obtaining the correct linear behavior---a 
massless spin two excitation, even with the correct coupling to matter---is 
not sufficient to show that one has a model with gravity.  Indeed, there 
are known examples (e.g., \cite{Zhang,Wen}) in which the correct nonlinear 
behavior seems to require a good deal of fine tuning.  

If one can obtain Lorentz invariance and diffeomorphism invariance,
however, these provide some very helpful constraints.  
As Kraichnan first showed \cite{Kraichnan}, if one starts with a massless
spin two field $h_{ab}$ on a manifold with a flat metric $\eta_{ab}$ and
assumes that its field equations can be derived from a Lorentz invariant,
diffeomorphism invariant action with no additional background structures, 
then the action can depend only on the combination $\eta_{ab}+h_{ab}$.  
This largely determines the form of the nonlinearities to be those of 
general relativity, thus fixing the self-coupling.

\subsection{Diffeomorphism invariance I}

Diffeomorphism invariance is a notoriously slippery concept in general 
relativity \cite{Anderson,Giulini}.  The rather heuristic form I will use
 is the absence of any nondynamical background structure that could 
define a preferred reference frame.  Most type I models have a nondynamical 
background, so the issue is again one of decoupling.  Most type II 
models do not, but one must show that the emergent spacetime is 
enough like a smooth manifold for diffeomorphism invariance to make 
sense at all.  Note that while diffeomorphism invariance and Lorentz
invariance are conceptually distinct, they are not completely unrelated:
in an ``already Lorentz invariant'' model, the only invariant background 
structures are a flat metric and a volume element, so the possible forms 
of diffeomorphism noninvariance are restricted.

As Witten has stressed \cite{Witten}, diffeomorphism invariance also
requires the absence of local observables \cite{Torre,Hartle}.  This 
presents yet another decoupling problem \cite{Elvang}: in type I models, 
all local observables, including any fundamental stress-energy tensor, 
must be invisible at the scale at which gravity emerges, while in type II 
models, the emergent spacetime should probably be free of local 
observables from the start.

For type I models, the problem of diffeomorphism invariance parallels 
the decoupling  problem for Lorentz invariance.  One case is known in 
which a weak form of diffeomorphism invariance appears \cite{Girellib}.
In this analog model, Nordstr{\o}m gravity emerges at lowest 
order,  but conformal invariance of the matter fields makes the flat 
background metric unobservable, leaving only a background conformal 
structure.  For other models, useful insights may come from the existing 
body of work on diffeomorphism invariance on a lattice.  While some 
of this work directly addresses emergent models \cite{Wetterich3}, 
much of it is in the context of lattice regularization of general 
relativity \cite{Dittrich0,Dittrich,Dittrich2,Kempf,Pullin,Pullin2}.  In 
particular, there are interesting ideas for obtaining an invariant lattice 
action---a ``perfect action''---from a Wilsonian coarse-graining of 
the continuum \cite{Dittrich0,Dittrich,Dittrich2,Kempf}, which could 
point to a new type of emergent model.  

A new problem arises in models in which the time evolution of the 
fundamental degrees of freedom depends on data in a finite region, 
as is the case in lattice models \cite{Wall}.  Consider two disjoint 
spatial regions $R_1$ and $R_2$ at time $t_1$, initially evolving 
independently, and let $I(R_1)$ and $I(R_2)$ be their respective
future domains of influence.  If these domains overlap
at some later time $t_2$, the relative rate of evolution can matter:
the data in $R_1$ at time $t_1$, for instance, will change the data 
in $I(R_1)\cap I(R_2)$ at time $t_2$, and thus the subsequent 
evolution of $R_2$.  Using a term from computer science, Wall 
calls this the ``race problem'': two independent regions are 
``racing'' toward the intersection $I(R_1)\cap I(R_2)$ , and the one 
that gets there first determines the subsequent evolution.  Such 
behavior, which is known to occur in particular models, clearly 
breaks diffeomorphism invariance; in an emergent gravity model, 
it would be interpreted as a failure of Hamiltonian constraints 
smeared by two different lapses to weakly commute.   Avoiding 
this problem seems very difficult, requiring either an arbitrary 
division of space into nonoverlapping regions or the imposition of 
extremely delicate consistency conditions.  It can be argued that 
these consistency conditions appear automatically for a ``perfect 
action,'' a discrete action that is already invariant under the full 
diffeomorphism group \cite{Dittrich0,Dittrich4}, but it is not  
clear how such a structure would arise from a more primitive
noninvariant theory.

For some type II models, another problem can occur.  While
such models involve emergent space, some (e.g., \cite{Marko,Lloyd})
include a time parameter to describe the evolution of the underlying 
degrees of freedom.  In such cases, one must worry about the
relationship between this fundamental time and the emergent time in
the description of gravity.  This is another decoupling problem:
the time in which the fundamental degrees of freedom evolve is a
background structure, and any coupling to the emergent degrees of
freedom would define a preferred time and break diffeomorphism 
invariance.

\subsection{Diffeomorphism invariance II}

Diffeomorphism invariance plays another key role in general relativity: 
it eliminates the spin zero and spin one degrees of freedom, leaving 
only spin two modes.  In the ADM formalism, the spatial metric 
and its conjugate momentum form six independent canonical pairs 
$(q_{ij},\pi^{ij})$, but the four diffeomorphism constraints 
eliminate four pairs.  It is crucial that the constraints are first class 
(i.e., that the commutator of two constraints is itself proportional to 
the constraints); a first class constraint eliminates two phase space
degrees of freedom, while a second class constraint eliminates only one
\cite{HenneauxTeitelboim}.  While the presence of additional degrees 
of freedom cannot be completely excluded by experiment, spin zero or 
one components of gravity are very strongly constrained, since, for 
example, they would imply violations of the principle of equivalence.

This aspect of diffeomorphism invariance presents a particularly strong
challenge for emergent models, since an ``approximate symmetry'' can
be qualitatively different from an exact one.  Suppose, for instance, that
an emergent weak gravitational field can be described at some length 
scale by a Lorentz invariant spin two field, but without the gauge
invariance corresponding to linearized diffeomorphisms.  The lowest order
action is then the Fierz-Pauli action for a field of mass $m$.
But although linear diffeomorphism invariance is restored in the 
$m\rightarrow0$ limit, that limit differs from weak field general
relativity \cite{VDV,Z}, and gives incorrect predictions for
Solar System tests.  This  van Dam-Veltman-Zakharov (vDVZ)  discontinuity 
arises because an extra scalar mode fails to decouple even in the 
massless limit.  Moreover, as noted earlier, nonlinear extensions of this 
model typically contain negative energy Boulware-Deser ghosts 
\cite{BD,Hinterbichler}.

As Vainshtein first pointed out \cite{Vainshtein}, the vDVZ discontinuity 
may indicate a breakdown of weak field perturbation theory: nonlinear 
effects proportional to inverse powers of $m$ may appear, signaling 
the onset of a strong coupling regime.  The unwanted scalar mode 
might then be screened, and a different perturbative expansion at 
short distances might more closely approximate general relativity 
(see \cite{Hinterbichler} for a review).  This mechanism has been 
confirmed for particular models (e.g., \cite{Babichev,Kaloper,DeFelice}), 
and with a very special choice of action, the Boulware-Deser ghosts 
may also be banished \cite{Hassan,deRham,RRosen}.  But if  the 
Vainshtein mechanism applies, it poses a new challenge: since the 
usual weak field approximation can no longer be trusted, one must 
work hard to find even a Newtonian limit for emergent gravity.

To a certain extent, this argument can be turned on its head: if one
is certain that a massless, Lorentz invariant, purely spin two field
has emerged with no lower spin partners, this strongly suggests the 
presence of diffeomorphism invariance or a similar symmetry.  A 
symmetric rank two tensor field $h_{ab}$ contains components of 
spin zero, one, and two, and the only known Lorentz covariant way to 
project out the lower spins is with a gauge invariance.  At linear 
order, the minimal requirement is invariance under ``transverse 
diffeomorphisms,'' diffeomorphisms generated by those vector fields 
$\xi^a$ for which $\partial_a\xi^a=0$ \cite{Bij,Alvarez,Blas}.  This 
group may be extended by including the remaining diffeomorphisms,
giving full diffeomorphism invariance, or by appending Weyl
transformations, yielding ``WTDiff'' invariance \cite{Alvarez}.
Contrary to popular folklore, however, neither extension is required
for a consistent nonlinear theory; Wald \cite{Wald} and Heiderich
and Unruh \cite{Heiderich} have explicitly constructed consistent
non-covariant models of interacting massless spin two fields,
and the latter contain explicit transverse diffeomorphism
invariance.\footnote{Wald notes that it may be difficult---in some
cases impossible---to couple such models to matter through the
standard stress-energy tensor.  But in the context of emergent
gravity, one cannot simply assume the standard coupling; one 
should start with the underlying theory and see what coupling 
emerges.}  

If one makes the much stronger assumption that the
field is sourced by its own stress-energy tensor, then, as noted
above, general relativity (and thus full diffeomorphism invariance)
will emerge \cite{Deser,Deser2}, although for the exceptional 
choice of a WTDiff-invariant linearized action one may instead
obtain unimodular gravity \cite{Alvarez,Blas,Alvarez2}.
The Weinberg-Witten theorem, described below, leads to a similar 
conclusion, that a pure spin two field with a conserved source must 
normally be a gauge theory.

If Lorentz invariance is not exact, the Fierz-Pauli action is no longer
unique.  The problem then becomes more difficult to analyze, although 
there are some models that appear to avoid both ghosts and the vDVZ 
discontinuity \cite{Pilo}.   But similar issues of extra ``gauge'' modes 
appear in other settings, such as lattice models \cite{Dittrich3}.   A 
basic lesson is that the linear approximation may be quite misleading; 
one must ensure that any undesirable lower spin modes decouple at the 
full nonlinear level.
 
\subsection{Diffeomorphism invariance and flat backgrounds}

It is worth noting a somewhat subtle technical issue in emergent 
diffeomorphism invariance.  Under an infinitesimal diffeomorphism 
generated by a vector field $\xi$, the metric transforms as
\begin{align}
g_{ab} \rightarrow g_{ab} + \nabla_a\xi_b + \nabla_b\xi_a
    = g_{ab} + g_{ac}\partial_b\xi^c + g_{bc}\partial_a\xi^c 
    + \xi^c\partial_c g_{ab}
\label{b61}
\end{align}
The last term is crucial: it reflects the fact that diffeomorphisms
``move points,'' and are not just ordinary pointwise gauge 
transformations.

Suppose, however, one expands around a flat metric $\eta_{ab}$.  
Then to lowest order, (\ref{b61}) becomes
\begin{align}
\eta_{ab} \rightarrow \eta_{ab} + \eta_{ac}\partial_b\xi^c 
    + \eta_{bc}\partial_a\xi^c  .
\label{b62}
\end{align}
The crucial derivative is now hidden.  This is not uncommon in
emergent gravity (for example, \cite{Wen,Xu}), where it may seem natural 
to build diffeomorphisms out of local gauge transformations.  But as a 
pointwise transformation, (\ref{b62}) is not yet a diffeomorphism---%
it is easy to check, for instance, that  the algebra of such transformations 
is not the algebra of diffeomorphisms---and the nonlinear interactions of
general relativity do not automatically appear.  As in the preceding 
section, a reliable demonstration of diffeomorphism invariance requires 
an expansion to nonlinear order.

\subsection{The Weinberg-Witten theorem}

In 1980, Weinberg and Witten proved a result that further constrains 
type I emergent models \cite{WW}.  The theorem can be stated as 
follows (see \cite{Jenkins,Loebbert} for further discussion):
\begin{center}
\parbox{.8\textwidth}{
Let $T^{\mu\nu}$ be a Lorentz covariant, conserved current.  Then no
massless spin two field can carry a nonzero charge under the operator
$P^\mu = \int d^3x\,T^{\mu 0}$.}
\end{center}
In particular, if $T^{\mu\nu}$ is a conserved stress-energy tensor, 
the theorem asserts that no massless ``graviton'' can carry energy or
momentum.

The Weinberg-Witten theorem uses no detailed properties beyond Lorentz
invariance and conservation, and applies to composite as 
well as elementary fields.  Naively, it would seem to rule out any theory,
including general relativity, in which the gravitational field carries
energy.  This cannot be the case, but by seeing how particular models 
evade the result, we can understand the true limits.

Let us start with general relativity in the weak field approximation.  
The obvious loophole is that the stress-energy tensor is not conserved, 
$\partial_\mu T^{\mu\nu}\ne0$, but only covariantly 
conserved, $\nabla_\mu T^{\mu\nu}=0$.  But one can always add
a gravitational stress-energy pseudotensor to form a conserved current.
Contrary to some claims in the literature, the resulting quantity can 
be fully Lorentz covariant: its definition requires a flat background 
metric, but this need not lead to any Lorentz violation 
\cite{Weinberg2}.  The result is not a tensor with respect to general 
coordinate transformations, but it is not obvious that this is relevant, 
since the proof of the theorem does not rely explicitly on general 
covariance.

The real issue is somewhat more subtle.  The Weinberg-Witten theorem
requires a ``pure'' spin two field, with no spin zero or spin one 
admixtures.  We may achieve this in two ways:
\begin{itemize}\renewcommand{\labelitemi}{\labelitemii}
\item We may project out the unphysical helicity zero and one states.
But such a projection is only Lorentz covariant up to a gauge 
transformation, violating one condition of the theorem.
\item We may appeal to diffeomorphism invariance to argue that 
the spin zero and one components of the metric are ``pure gauge,'' 
and therefore irrelevant.  But this argument only works in a gauge
covariant formulation.  We are caught: the stress-energy pseudotensor 
is not covariant, while the covariant stress-energy tensor is not 
conserved.
\end{itemize}
It is this interaction of gauge invariance and Lorentz invariance that
provides the loophole.\footnote{This explains the apparent conflict
between the Weinberg-Witten theorem and the soft graviton theorem of
section \ref{equiv}.  The soft graviton theorem is an on-shell result,
requiring only Lorentz invariance of the S-matrix; it has no
requirement of a local, conserved Lorentz covariant stress-energy tensor.}

Now, if a model of emergent gravity reproduces general relativity above 
some length scale $L$, the same loophole should apply at that scale.  
The question becomes whether the Weinberg-Witten theorem restricts 
the model at shorter scales.  Possible solutions include 
\cite{Sindonirev,Jenkins2}
\begin{enumerate}
\item Broken Lorentz invariance: in analog models \cite{LV} and models
in which the graviton is a Goldstone boson for broken Lorentz invariance
\cite{Kraus,Sindonirev,Jenkins}, for instance, the fundamental degrees
of freedom are not Lorentz invariant, evading one condition of the theorem.
\item Nonlocality: in Sundrum's ``fat graviton'' model \cite{Sundrum}, 
and arguably the AdS/CFT correspondence, gravitons are nonlocal, and 
do not couple to a local stress-energy tensor.
\item No spin two fields below $L$: if spin two fields first emerge 
at the same scale as general relativity, there is no room for the 
Weinberg-Witten theorem to apply.  For example, in models in which 
the background manifold is topological \cite{Amati,Floreanini,Wetterich,%
Wetterich2,Alfaro}, there may be no nontrivial conserved stress-energy 
tensor at all at small scales.
\item Emergent spacetime: in type II models of emergence, the basic
setting of the Weinberg-Witten theorem, spin two excitations in a flat
spacetime, is absent at the fundamental level, though one must check
carefully at larger scales.
\end{enumerate}

\subsection{Where does the emergent theory live?}

In section \ref{cats}, I introduced two general categories of emergent 
models: type I models, which assume a background ``environment,'' 
and type II models, in which spacetime itself is emergent.  In some 
ways, type II models are more appealing: if macroscopic gravity is a 
characteristic of the structure of spacetime, shouldn't the structure 
emerge with the spacetime itself?  But for the same reason, type II 
models are also much harder to connect to known physics.

Most type I models, on the other hand, present us with a basic question:
what determines the environment?  For example, in
models of gravitons as composite spin two particles in a flat Minkowski
space, why is the background spacetime flat?  A century ago, this could 
have passed as an ``obvious'' assumption.  But once we know that curvature
can be dynamical, we cannot simply forget that knowledge; we now know
that we are secretly postulating a field equation, $R_{abcd}=0$, for the
background.  Similarly, for models on a fixed lattice, what fixes the 
lattice topology and spacing?  We know that these features can be 
dynamical, as they are in Regge calculus; why do they not evolve in these 
models?  In particular, what prevents the back-reaction of the emergent 
gravitational degrees of freedom on the ``fixed'' components of the 
environment?

These are not experimental questions, and the answer could be
simply, ``That's the way Nature is.''  But the spirit of the emergent 
gravity program is to replace general relativity with something more
fundamental, and a fixed background seems to be a
step in the wrong direction.  One interesting attempt to address such 
questions comes from work on ``noiseless subsystems'' \cite{Marko4}, 
in which the emergent structure is \emph{defined} by its decoupling 
from the background, but it remains to be seen whether such a special
characteristic can hold for realistic models.

\subsection{The usual problems of quantum gravity}

Emergent gravity is sometimes advertised as a solution to the problems 
of quantizing general relativity.  This is not an unreasonable hope: the
underlying degrees of freedom may be renormalizable, for instance, or 
may have a discrete structure that provides a natural cutoff.  But there 
is more to quantum gravity than renormalizability, and it is not clear 
that emergent models can do better than ordinary general relativity 
in addressing fundamental conceptual problems \cite{Carlip}.

For example, a quantum theory of general relativity has no local
observables \cite{Torre,Hartle}, and it is quite difficult to reconstruct a 
local picture of physics.  As long as an emergent model recovers 
diffeomorphism invariance, this problem will persist, at least at the 
scale at which an effective gravitational description is possible.  

Similarly, many aspects of the infamous ``problem of time'' \cite{Kuchar}
will remain.  A model that recovers diffeomorphism invariance will
have no preferred time coordinate, and will have a
Hamiltonian constraint rather than a Hamiltonian.  If, on the other hand,
the fundamental degrees of freedom have a preferred time that does not
completely decouple from the gravitational degrees of freedom, the absence
of a Hamiltonian constraint will lead to extraneous degrees of freedom, 
with the concomitant problems discussed in section 3.5.  Even in this 
case, the Weinberg-Witten theorem will imply the absence of a Lorentz 
invariant, conserved Hamiltonian at the scale at which gravity emerges.

Causality is problematic as well.  A fundamental feature of ordinary 
quantum field theory is that spacelike separated operators commute.
But if the metric---even an emergent one---is subject to quantum 
fluctuations, there will be no fixed light cones to define spacelike and
timelike separation.  In some type I models, one might hope that the
underlying ``environment'' defines an absolute causality.  This would
be radically different from the analogous classical situation, however.   
Classically, if one treats the gravitational field as a massless spin two 
field $h_{\mu\nu}$ propagating on a flat background, one finds that the
nonlinearities of the action force matter to couple to the full metric
$g_{\mu\nu} = \eta_{\mu\nu} + h_{\mu\nu}$, hiding all traces of the 
flat background metric \cite{Kraichnan,Deser}.  In particular, the support 
of Greens functions lies within the $g$ light cones, not the $\eta$ light 
cones.  It is true that the $g$ light cones normally lie inside the $\eta$ 
light cones \cite{Visser}, so the ``emergent'' metric does not violate 
background causality.  But this result depends on special features
of classical general relativity, and even there it holds only if matter 
satisfies the null energy condition, a condition that quantum 
fluctuations do not obey \cite{Ford}.    

\section{Where we stand}

Einstein gravity is a robust theory, which can be reached from many 
different starting points.  Chapter 17 of the famous textbook by Misner, 
Thorne, and Wheeler describes six routes to the Einstein field equations 
\cite{MTW}; the EPS derivation described in section \ref{EPSsec} 
provides a seventh.   This might offer hope that emergent gravity could 
also lead to the same large scale physics.

As I have tried to show, life is not so easy.  Gravity may be an
emergent phenomenon, but models of emergent gravity faces formidable
obstacles.  For all its simplicity, general relativity rests heavily on a few 
fundamental features---local Lorentz invariance, the principle of 
equivalence, diffeomorphism invariance and background independence---%
that are not easy to mock up.  

Moreover, these features are intertwined.  A local model with a 
background time, for instance, must lose all traces of its Hamiltonian 
at the scale at which gravity emerges, or the Weinberg-Witten theorem 
might force the emergent theory to be Lorentz-violating.  A model 
in which the gravitational field does not couple universally to matter is 
likely to have no single conserved stress-energy tensor, and thus no 
suitable gauge invariance for a spin two field.  A model whose background 
environment fails to sufficiently decouple will have problems not 
only with Lorentz invariance, but with diffeomorphism invariance,
the principle of equivalence, and, quite likely, Boulware-Deser ghosts.

These difficulties do not mean that the search for emergent gravity is 
doomed.  But they suggest that current ad hoc approaches  are unlikely 
to succeed.  While there is much to be learned from such models, 
it seems likely that a successful theory of emergent gravity will require 
some more fundamental principle, as yet unknown, to allow its emergent 
properties to be organized into a realistic model of spacetime.

\vspace{1.5ex}
\begin{flushleft}
\large\bf Acknowledgments
\end{flushleft}

Much of this work grew out of the September 2011 ``Gravity as 
Thermodynamics'' workshop at SISSA, Trieste and a series of 
Peyresq Physics conferences.  I thank the organizers and participants 
for pushing me to think harder about these topics.  I would 
particularly like to thank  Stanley Deser, Bianca Dittrich, Stefano 
Liberati, Thanu Padmanabhan, and Dean Rickles for useful suggestions.  
The Trieste workshop was supported by the European Science 
Foundation, and Peyresq Physics by OLAM Association pour la 
Recherche Fundamentale, Bruxelles.  My own work was supported 
in part by US Department of Energy grant DE-FG02-91ER40674.

\end{document}